\documentclass{article}
\usepackage{ijcai24}

\usepackage{times}
\usepackage{soul}
\usepackage{url}
\usepackage[hidelinks]{hyperref}
\usepackage[utf8]{inputenc}
\usepackage[small]{caption}
\usepackage{graphicx}
\usepackage{amsmath}
\usepackage{amsthm}
\usepackage{booktabs}
\usepackage{algorithm}
\usepackage{algorithmic}
\usepackage[switch]{lineno}
\usepackage{amsfonts}
\usepackage{arydshln}
\usepackage{xcolor}
\usepackage{color}
\usepackage{tikz}
\usepackage{pgfplots}
\pgfplotsset{compat=1.18}
\usepackage{subfigure}
\usepackage{makecell}
\usepackage{multirow}
\usepackage{threeparttable}
\usepackage{hyperref}
\newcommand{\citet}[1]{\citeauthor{#1} \shortcite{#1}}

\title{Recent Advances in End-to-End Simultaneous Speech Translation}

\author{
Xiaoqian Liu$^1$\and
Guoqiang Hu$^2$\and
Yangfan Du$^1$\and
Erfeng He$^1$\and \\
Yingfeng Luo$^1$ \and
Chen Xu$^3$ \and
Tong Xiao$^{1,4}$ \textnormal{and}
Jingbo Zhu$^{1,4}$\footnote{Corresponding author.}\\
\affiliations
$^1$School of Computer Science and Engineering, Northeastern University, Shenyang, China\\
$^2$International School, Jinan University, Guangzhou, China\\
$^3$College of Computer Science and Technology, Harbin Engineering University, Harbin, China\\
$^4$NiuTrans Research, Shenyang, China\\
\emails
\{liuxiaoqian0319, luoyf98, xuchennlp, giannis-huguogiang\}@outlook.com,\\
\{dduyangfan, heerfeng1023\}@gmail.com, 
\{xiaotong, zhujingbo\}@mail.neu.edu.cn
}

\begin{document}

\maketitle

\begin{abstract}
Simultaneous speech translation (SimulST) is a demanding task that involves generating translations in real-time while continuously processing speech input. This paper offers a comprehensive overview of the recent developments in SimulST research, focusing on four major challenges. Firstly, the complexities associated with processing lengthy and continuous speech streams pose significant hurdles. Secondly, satisfying real-time requirements presents inherent difficulties due to the need for immediate translation output. Thirdly, striking a balance between translation quality and latency constraints remains a critical challenge. Finally, the scarcity of annotated data adds another layer of complexity to the task. Through our exploration of these challenges and the proposed solutions, we aim to provide valuable insights into the current landscape of SimulST research and suggest promising directions for future exploration.

\end{abstract}

\section{Introduction}
\label{intro}
End-to-end simultaneous speech translation (SimulST) is a task of generating translation text in the target language while receiving speech input in the source language \cite{SimulMTtoSimulST}. This process of directly processing input and providing translation is seamless and continuous, giving a faster and more natural translation experience. It is especially beneficial for real-time conversations, voice conferencing, and other scenarios that require fast and smooth communication, and therefore has received widespread attention and progress in recent years \cite{AlignAtt}. Meanwhile, since speech translation (ST) itself is already a cross-lingual and cross-modal task \cite{xusurvey}, the demand for streaming generation makes it even more complex. 

\input{fig/fig1}

Figure \ref{overview} presents an overview of the SimulST model. Based on the encoder-decoder structure, the model also needs an additional segmentation module and a simultaneous read-write module for streaming inference. Giving the training data as a triple $\mathcal{D}=(S, X, Y)$, $S$ denotes the acoustic features extracted from the input speech in the source language, $X$ denotes the corresponding transcription, and $Y$ denotes the text in the target language. Considering a segmentation strategy to obtain input units and a simul R-W policy $g$, we denote the number of speech features as $g(t)$ at the translation of the $t$-th token $y_t$. The training objective using cross-entropy loss parameterized by $\theta$ can be formalized as:
\begin{eqnarray}
\mathcal{L}_{\theta}  = -\mathbb{E}_{( \hat{s},y_{<t})}\sum_{t=1}^T \log \textrm{P} (y_t | y_{<t}, \hat{s}; \theta)
\end{eqnarray}
where ${\hat{s}}$ denotes the current received features $[s_1  ...  s_{ g(t)}]$. The determination of $g(t)$ is crucial, demonstrating the important role of streaming input processing and decision-making methods. Moreover, the latency cannot be directly optimized through the traditional loss $\mathcal{L_{\theta}}$, which reflects the complexity of SimulST. Hence, existing research primarily focuses on addressing the following key challenges and issues:

\definecolor{qblue}{RGB}{68,114,196}
\definecolor{qpurple}{RGB}{187,161,203}
\definecolor{qgreen}{RGB}{112,173,71}
\definecolor{qorange}{RGB}{237,125,49}
\definecolor{qpink}{RGB}{225,97,210}
\definecolor{qred}{RGB}{255,99,71}
\definecolor{qyellow}{RGB}{255,192,0}
\definecolor{sgreen}{RGB}{197,224,180}
\definecolor{syellow}{RGB}{255,230,153}
\definecolor{sblack}{RGB}{242,242,242}
\definecolor{backyellow}{RGB}{251,232,218}
\definecolor{backgreen}{RGB}{229,241,221}
\definecolor{backblue}{RGB}{189,215,238}
\begin{figure*}
    \centering
    \tikzstyle{textonly} = [font=\scriptsize,align=center]
    \tikzstyle{sublayer} = [rectangle,draw,minimum width=2.5cm,rounded corners=2.5pt,align=left,inner sep=1pt,minimum height=0.4cm,font=\scriptsize]
   \begin{tikzpicture}
    \node[sublayer,fill=white!70,draw=black,thick](simulst) at (0,0) {SimulST};
    \node[sublayer,fill=qorange!30,draw=qorange,thick,align=center,minimum width=3.3cm](segment) at ([yshift=3.3cm,xshift=2.3cm]simulst.east) {Segmentation Stategy};
    \node[sublayer,fill=qorange!30,draw=qorange,thick](fixed-length) at ([yshift=0.55cm,xshift=2cm]segment.east) {Fixed-length};
    \node[sublayer,fill=qorange!30,draw=qorange,thick](wordbd) at ([yshift=0cm,xshift=2cm]segment.east) {Word-based};
    \node[sublayer,fill=qorange!30,draw=qorange,thick](adaseg) at ([yshift=-0.55cm,xshift=2cm]segment.east) {Adaptive Segmentation};
    \node[sublayer,minimum width=4.3cm,dashed,draw=qorange,thick,fill=qorange!10]at([xshift=2.55cm]wordbd.east){Processing long-form speech inputs};
    \node[sublayer,fill=qgreen!30,draw=qgreen,thick,minimum width=3.3cm](rwpolicy) at ([yshift=0.75cm,xshift=2.3cm]simulst.east) {Simul Read-Write Policy};
    \node[sublayer,fill=qgreen!30,draw=qgreen,thick](rwfixed) at ([yshift=0.85cm,xshift=2cm]rwpolicy.east) {Fixed};
    \node[sublayer,fill=qgreen!30,draw=qgreen,thick](waitk) at ([xshift=2cm]rwfixed.east) {Wait-\textit{k} Stride-\textit{n}} ;
    \node[sublayer,fill=qgreen!30,draw=qgreen,thick](waitkstride) at ([xshift=2cm,yshift=-0.55cm]rwfixed.east) {Wait-(\textit{k, s, n})};
    \node[sublayer,fill=qgreen!30,draw=qgreen,thick](waitksn) at ([xshift=2cm,yshift=0.55cm]rwfixed.east) {Wait-\textit{k}};
    \node[sublayer,fill=qgreen!30,draw=qgreen,thick](flexible) at ([yshift=-0.85cm,xshift=2cm]rwpolicy.east) {Flexible};      
    \node[sublayer,fill=qgreen!30,draw=qgreen,thick](aed) at ([yshift=0.55cm,xshift=2cm]flexible.east) {Attn-based Enc-Dec};  
    \node[sublayer,fill=qgreen!30,draw=qgreen,thick](o2o) at ([yshift=-0.55cm,xshift=2cm]flexible.east) {Offline-to-Simul};  
    \node[sublayer,fill=qgreen!30,draw=qgreen,thick](trans) at ([xshift=2cm]flexible.east) {Transducer};
    \node[sublayer,minimum width=4cm,dashed,draw=qgreen,thick,fill=qgreen!10]at([xshift=2.3cm,yshift=0.45cm]aed.east){Satisfying real-time requirements};
    \node[sublayer,fill=qyellow!30,draw=qyellow,thick,minimum width=3.3cm](ql) at ([yshift=-1.55cm,xshift=2.3cm]simulst.east) {Evaluation Metrics};
    \node[sublayer,fill=qyellow!30,draw=qyellow,thick](eval) at ([yshift=0.3cm,xshift=2cm]ql.east) {Quality-based Metrics};
    \node[sublayer,fill=qyellow!30,draw=qyellow,thick](trade) at ([yshift=-0.3cm,xshift=2cm]ql.east) {Latency-based Metrics};
    \node[sublayer,minimum width=4.3cm,dashed,draw=qyellow,thick,fill=qyellow!10]at([xshift=2.55cm,yshift=-0.25cm]eval.east){Balancing quality and latency trade-offs};
    \node[sublayer,fill=qblue!30,draw=qblue,thick,minimum width=3.3cm](ds) at ([yshift=-2.72cm,xshift=2.3cm]simulst.east) {Augmented Training Methods};
    \node[sublayer,fill=qblue!30,draw=qblue,thick](mtasks) at ([yshift=0.3cm,xshift=2cm]ds.east) {Data Augmentation};
    \node[sublayer,fill=qblue!30,draw=qblue,thick](da) at ([yshift=-0.3cm,xshift=2cm]ds.east) {Multi-task Learning};
    \node[sublayer,minimum width=4.3cm,dashed,draw=qblue,thick,fill=qblue!10]at([xshift=2.55cm,yshift=-0.25cm]mtasks.east){Addressing data scarcity issues};

\draw[-,thick](simulst.east)--([xshift=0.375cm]simulst.east)--([yshift=3.3cm,xshift=0.375cm]simulst.east)--(segment.west);
\draw[-,thick](simulst.east)--([xshift=0.375cm]simulst.east)--([yshift=0.75cm,xshift=0.375cm]simulst.east)--(rwpolicy.west);
\draw[-,thick](simulst.east)--([xshift=0.375cm]simulst.east)--([yshift=-1.55cm,xshift=0.375cm]simulst.east)--(ql.west);
\draw[-,thick](simulst.east)--([xshift=0.375cm]simulst.east)--([yshift=-2.72cm,xshift=0.375cm]simulst.east)--(ds.west);

\draw[-,thick](segment.east)--([xshift=0.375cm]segment.east)--([yshift=0.55cm,xshift=0.375cm]segment.east)--(fixed-length.west);
\draw[-,thick](segment.east)--([xshift=0.375cm]segment.east)--([yshift=-0.55cm,xshift=0.375cm]segment.east)--(adaseg.west);
\draw[-,thick](segment.east)--(wordbd.west);

\draw[-,thick](rwpolicy.east)--([xshift=0.375cm]rwpolicy.east)--([yshift=0.85cm,xshift=0.37cm]rwpolicy.east)--(rwfixed.west);
\draw[-,thick](rwfixed.east)--(waitk.west);
\draw[-,thick](rwfixed.east)--([xshift=0.375cm]rwfixed.east)--([yshift=-0.55cm,xshift=0.37cm]rwfixed.east)--(waitkstride.west);
\draw[-,thick](rwfixed.east)--([xshift=0.375cm]rwfixed.east)--([yshift=0.55cm,xshift=0.37cm]rwfixed.east)--(waitksn.west);

\draw[-,thick](rwpolicy.east)--([xshift=0.375cm]rwpolicy.east)--([yshift=-0.85cm,xshift=0.37cm]rwpolicy.east)--(flexible.west);
\draw[-,thick](flexible.east)--([xshift=0.375cm]flexible.east)--([yshift=0.55cm,xshift=0.375cm]flexible.east)--(aed.west);
\draw[-,thick](flexible.east)--(trans.west);
\draw[-,thick](flexible.east)--([xshift=0.375cm]flexible.east)--([yshift=-0.55cm,xshift=0.375cm]flexible.east)--(o2o.west);

\draw[-,thick](ql.east)--([xshift=0.375cm]ql.east)--([yshift=0.3cm,xshift=0.375cm]ql.east)--(eval.west);
\draw[-,thick](ql.east)--([xshift=0.375cm]ql.east)--([yshift=-0.3cm,xshift=0.375cm]ql.east)--(trade.west);

\draw[-,thick](ds.east)--([xshift=0.375cm]ds.east)--([yshift=0.3cm,xshift=0.375cm]ds.east)--(mtasks.west);
\draw[-,thick](ds.east)--([xshift=0.375cm]ds.east)--([yshift=-0.3cm,xshift=0.375cm]ds.east)--(da.west);

    
\end{tikzpicture}
\caption{Key challenges to address in the task of SimulST and their corresponding solutions.}
\label{taxonomy}
\end{figure*}
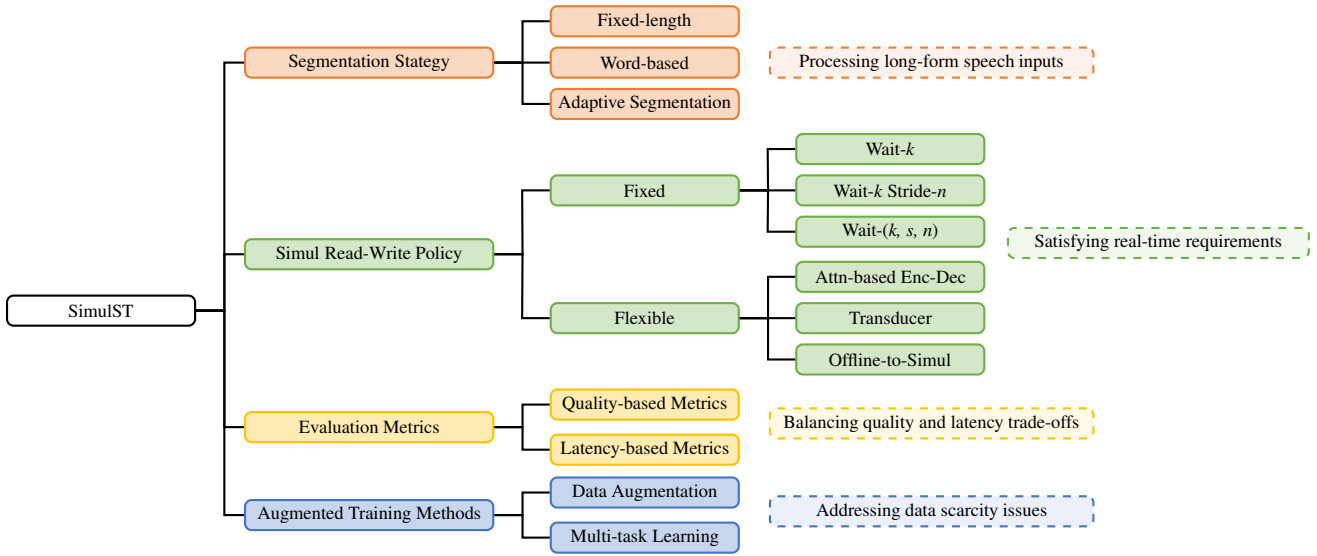

\begin{itemize}
    \item \textbf{Processing long-form inputs.} SimulST demands models to possess both translation accuracy and low-latency capabilities. However, lengthy and continuous inputs fail to meet the low-latency requirements for real-time output \cite{Zhangacl}.
    \item \textbf{Satisfying real-time requirements.} To the current input segment, the model needs to make decisions regarding whether to generate a new translation \cite{SimulMTtoSimulST}. Premature output may result in incomplete information, leading to poorer translations. Conversely, delaying output may introduce high latency, thereby impacting user experience.
    \item \textbf{Balancing quality and latency trade-offs.} There is no single evaluation metric that can simultaneously address both quality and latency \cite{atd}. Achieving a balance between quality and latency is indeed difficult, especially in the context of SimulST.
    \item \textbf{Addressing data scarcity issues.} Unlike related fields such as automatic speech recognition (ASR) and machine translation (MT) which have abundant training data \cite{Tagged}, SimulST suffers from a scarcity of annotated data, which exacerbates the complexity as the models are hard to be adequately trained.
\end{itemize}

These factors collectively contribute to the intricate nature of the SimulST task. Existing studies have proposed solutions to these challenges, but there is currently a lack of a comprehensive overview to thoroughly summarize their practices. We find that there are some previous related studies, in which \citet{xusurvey} aim at offline ST tasks, \citet{longformsurvey} faces long-form inputs, and \citet{survey23} summarize the whole speech-to-text technology. Our work is different from theirs since we give a more complete and comprehensive introduction to SimulST. 

As shown in Figure \ref{overview} and Figure \ref{taxonomy}, we structure the paper as follows. Section \ref{seg} introduces \textit{\textbf{segmentation strategies}} for (a) processing continuous speech inputs, realizing \textit{where} to segment the input into a suitable acoustic unit. Methods can be divided into fixed-length, word-based, and adaptive segmentation methods. Section \ref{sec:3} describes the \textit{\textbf{simultaneous read-write policies}} including fixed read-write methods and flexible ones to judge \textit{when} to output according to the currently obtained units, satisfying (b) real-time requirements. Section \ref{sec:4} introduces the studies related to (c) quality and latency which puts forward two kinds of \textit{\textbf{metrics}} to ensure a comprehensive \textit{\textbf{evaluation}}. Section \ref{sec:5} describes studies of the \textit{\textbf{augmented training methods}} including data enhancement and multi-task learning to tackle the (d) data scarcity issues. Finally, Section \ref{sec:6} anticipates some promising directions for \textit{\textbf{future SimulST research}}, including studies in multilingual SimulST and combining them with large language models.

\section{Segmentation Strategies}
\label{seg}

SimulST needs to read long-form streaming inputs and generate partial translation at inference time. Hence, it is crucial to select appropriate segmentation strategies to furnish the model with suitable units during inference. However, speech is a continuous sequence, and the absence of distinct boundaries poses a great challenge in achieving accurate segmentation. The segmentation strategies encompass three specific methods. In Section \ref{sec:2.1}, we elucidate the fixed-length strategies,  followed by an exposition of the word-based strategies in Section \ref{sec:2.2}. Moving to Section \ref{sec:2.3}, we present the adaptive segmentation strategies. 

\subsection{Fixed-length Strategies}
\label{sec:2.1}
Fixed-length strategies represent one of the simplest segmentation approaches. As illustrated in Figure \ref{segs} (a), it assumes that a certain number of speech frames (e.g. 280ms or 400 frames) equate to a fixed count of words, dividing the speech into equally-sized segments with a consistent frame length \cite{waitkstriden,SH}. This method bears resemblance to incremental encoding as discussed in \cite{waitk}, with each segment akin to a token.

The basic SimulST model is based on vanilla Transformer \cite{transformer} in Figure \ref{models} (a), namely a segment-based Transformer, in which the self-attention module attends to the entire sequence and limits the streaming capability of the model. Building upon this issue, \citet{augmemory} introduce an augmented memory encoder to divide the input sequence into sub-utterance level segments where each overlaps with previous and subsequent ones to capture left and right contexts. By computing self-attention only within each segment, the model can handle long input sequences while significantly reducing complexity. To enhance computational efficiency, \citet{ImplicitMemory} improve the model with implicit memory, which captures the representation of previous segments during encoding and implicitly preserves based on attention, thus eliminating the need for a memory bank. Due to discrepancies between training and inference in segment-based Transformer models, they subsequently propose a shiftable context method in \cite{Shiftable} to produce consistent segment sizes for better alignment.

\subsection{Word-based Strategies}
\label{sec:2.2}
While studies based on fixed-length segmentation methods have achieved numerous successes, making decisions on every fixed number of frames often leads to suboptimal results. This is because the segment boundaries may not align with the natural endings of pronunciations, thus disrupting acoustic integrity. To alleviate this situation, several unfixed-length strategies have been proposed, Among them, the word-based strategies expect to determine segmentation by aligning with corresponding words \cite{SimulMTtoSimulST}, as illustrated in Figure \ref{segs} \textit{(b)}. 

Word-based strategies typically involve introducing additional detectors or similar modules to detect boundaries, representing a hard alignment approach. Meanwhile, Connectionist Temporal Classification (CTC) \cite{ctc} proves effective in detecting word boundaries and is widely used when mapping frame-level classification outputs of speech sequences to text sequences. So in \cite{SimulSpeech}, a speech segmenter is added after the encoder to detect word boundaries and segment the input streaming speech using CTC loss. Besides, they introduce two knowledge distillation methods to ensure the performance. To relieve the burden of the encoder, \citet{realtrans} decouple it into three parts. They weight and aggregate the detected frames by a CTC module and introduce a blank penalty for non-blank labels. In addition, \citet{impact} leverage an additional toolkit with oracle word boundaries to segment input into corresponding words. In a word, these studies employ external segmentation modules, always leaving a gap between the segmentation and translation model.

\input{fig/fig3}
\subsection{Adaptive Segmentation Strategies}
\label{sec:2.3}
Different from previous systems that treat speech with fixed time-span as an acoustic unit or attempt to locate word boundaries, adaptive segmentation strategies depicted in Figure \ref{segs} \textit{(c)} detect boundaries of proper speech units. These strategies consider more meaningful information, such as semantic consistency, or incorporate input segmentation into the model training process.

To realize adaptive segmentation, \citet{whentotrans} propose MoSST, introducing a Monotonic Segmentation Module (MSM) to handle streaming speech input. The MSM dynamically reads the acoustic information from the encoder and locates the boundaries of meaningful speech units instead of segmentation. Inspired by the integrate-and-fire (IF) model, it incrementally integrates input when the information is not enough; once sufficient, it enters a firing mode, during which a new token is generated.

Another implementation is based on the concept of Meaningful Units (MUs), defined as the minimum speech segments whose translation will not be altered by subsequent speech \cite{adaptiveseg}. Therefore, a detection module is designed to dynamically detect MUs by comparing the translation of every speech prefix segment with the full-speech translation. Once an MU is detected, it is fed into the model for inference.

Different from adding a heuristic detector, \citet{Zhangacl} propose Differentiable Segmentation (DiSeg), which predicts a Bernoulli variable $0/1$ for each speech feature to indicate when to segment streaming speech inputs. It can be trained jointly with the SimulST model, allowing segmentation to be integrated into the translation process. Since DiSeg learns segmentation directly, it can handle simultaneous and offline speech translation with a unified model.

To summarize, as shown in Figure \ref{overview} \textit{(a)}, some studies introduce segmentation strategies before the encoder, while others opt for segmentation after encoding. However, regardless of their placement, these modules transform the original continuous inputs into reasonable acoustic units for SimulST.

\definecolor{qblue}{RGB}{68,114,196}
\definecolor{qpurple}{RGB}{187,161,203}
\definecolor{qgreen}{RGB}{112,173,71}
\definecolor{qorange}{RGB}{237,125,49}
\definecolor{qpink}{RGB}{249,199,207}
\definecolor{qred}{RGB}{255,99,71}
\definecolor{qyellow}{RGB}{255,192,0}
\definecolor{sgreen}{RGB}{197,224,180}
\definecolor{syellow}{RGB}{255,230,153}
\definecolor{sblack}{RGB}{242,242,242}
\definecolor{backyellow}{RGB}{251,232,218}
\definecolor{backgreen}{RGB}{229,241,221}
\definecolor{backblue}{RGB}{189,215,238}
\begin{figure}
\centering
\begin{tikzpicture}

\tikzstyle{textonly} = [font=\scriptsize,align=left];
\tikzstyle{squrd} = [rectangle,draw,minimum width=0.79cm,rounded corners=0pt,align=center,inner sep=3pt,minimum height=0.79cm,font=\scriptsize];

\node[textonly](originl) at (-10,0) {};
\node[squrd,draw opacity=0.5](a11)at(originl){};
\node[squrd,draw opacity=0.5](a12)at([xshift=0.395cm]a11.east){};
\node[squrd,draw opacity=0.5](a13)at([xshift=0.395cm]a12.east){};
\node[squrd,draw opacity=0.5](a14)at([xshift=0.395cm]a13.east){};
\node[squrd,draw opacity=0.5](a15)at([xshift=0.395cm]a14.east){};
\node[squrd,draw opacity=0.5](a16)at([xshift=0.395cm]a15.east){};
\node[squrd,draw opacity=0.5](a17)at([xshift=0.395cm]a16.east){};

\node[squrd,draw opacity=0.5](a21)at([yshift=-0.395cm]a11.south){};
\node[squrd,draw opacity=0.5](a22)at([xshift=0.395cm]a21.east){};
\node[squrd,draw opacity=0.5](a23)at([xshift=0.395cm]a22.east){};
\node[squrd,draw opacity=0.5](a24)at([xshift=0.395cm]a23.east){};
\node[squrd,draw opacity=0.5](a25)at([xshift=0.395cm]a24.east){};
\node[squrd,draw opacity=0.5](a26)at([xshift=0.395cm]a25.east){};
\node[squrd,draw opacity=0.5](a27)at([xshift=0.395cm]a26.east){};

\node[squrd,draw opacity=0.5](a31)at([yshift=-0.395cm]a21.south){};
\node[squrd,draw opacity=0.5](a32)at([xshift=0.395cm]a31.east){};
\node[squrd,draw opacity=0.5](a33)at([xshift=0.395cm]a32.east){};
\node[squrd,draw opacity=0.5](a34)at([xshift=0.395cm]a33.east){};
\node[squrd,draw opacity=0.5](a35)at([xshift=0.395cm]a34.east){};
\node[squrd,draw opacity=0.5](a36)at([xshift=0.395cm]a35.east){};
\node[squrd,draw opacity=0.5](a37)at([xshift=0.395cm]a36.east){};

\node[squrd,draw opacity=0.5](a41)at([yshift=-0.395cm]a31.south){};
\node[squrd,draw opacity=0.5](a42)at([xshift=0.395cm]a41.east){};
\node[squrd,draw opacity=0.5](a43)at([xshift=0.395cm]a42.east){};
\node[squrd,draw opacity=0.5](a44)at([xshift=0.395cm]a43.east){};
\node[squrd,draw opacity=0.5](a45)at([xshift=0.395cm]a44.east){};
\node[squrd,draw opacity=0.5](a46)at([xshift=0.395cm]a45.east){};
\node[squrd,draw opacity=0.5](a47)at([xshift=0.395cm]a46.east){};

\node[squrd,draw opacity=0.5](a51)at([yshift=-0.395cm]a41.south){};
\node[squrd,draw opacity=0.5](a52)at([xshift=0.395cm]a51.east){};
\node[squrd,draw opacity=0.5](a53)at([xshift=0.395cm]a52.east){};
\node[squrd,draw opacity=0.5](a54)at([xshift=0.395cm]a53.east){};
\node[squrd,draw opacity=0.5](a55)at([xshift=0.395cm]a54.east){};
\node[squrd,draw opacity=0.5](a56)at([xshift=0.395cm]a55.east){};
\node[squrd,draw opacity=0.5](a57)at([xshift=0.395cm]a56.east){};

\node[squrd,draw opacity=0.5](a61)at([yshift=-0.395cm]a51.south){};
\node[squrd,draw opacity=0.5](a62)at([xshift=0.395cm]a61.east){};
\node[squrd,draw opacity=0.5](a63)at([xshift=0.395cm]a62.east){};
\node[squrd,draw opacity=0.5](a64)at([xshift=0.395cm]a63.east){};
\node[squrd,draw opacity=0.5](a65)at([xshift=0.395cm]a64.east){};
\node[squrd,draw opacity=0.5](a66)at([xshift=0.395cm]a65.east){};
\node[squrd,draw opacity=0.5](a67)at([xshift=0.395cm]a66.east){};

\node[squrd,draw opacity=0.5](a71)at([yshift=-0.395cm]a61.south){};
\node[squrd,draw opacity=0.5](a72)at([xshift=0.395cm]a71.east){};
\node[squrd,draw opacity=0.5](a73)at([xshift=0.395cm]a72.east){};
\node[squrd,draw opacity=0.5](a74)at([xshift=0.395cm]a73.east){};
\node[squrd,draw opacity=0.5](a75)at([xshift=0.395cm]a74.east){};
\node[squrd,draw opacity=0.5](a76)at([xshift=0.395cm]a75.east){};
\node[squrd,draw opacity=0.5](a77)at([xshift=0.395cm]a76.east){};

\draw[->,line width=2pt,qorange]([yshift=0.395cm]a11.west)--([yshift=0.395cm]a13.west);
\draw[->,line width=2pt,qblue]([yshift=0.395cm]a13.west)--([yshift=-0.395cm]a13.west);
\draw[->,line width=2pt,qorange]([xshift=-0.395cm]a13.south)--([xshift=0.395cm]a13.south);
\draw[->,line width=2pt,qblue]([yshift=0.395cm]a24.west)--([yshift=-0.395cm]a24.west);
\draw[->,line width=2pt,qorange]([xshift=-0.395cm]a24.south)--([xshift=0.395cm]a24.south);
\draw[->,line width=2pt,qblue]([yshift=0.395cm]a35.west)--([yshift=-0.395cm]a35.west);
\draw[->,line width=2pt,qorange]([xshift=-0.395cm]a35.south)--([xshift=0.395cm]a35.south);
\draw[->,line width=2pt,qblue]([yshift=0.395cm]a46.west)--([yshift=-0.395cm]a46.west);
\draw[->,line width=2pt,qorange]([xshift=-0.395cm]a46.south)--([xshift=0.395cm]a46.south);
\draw[->,line width=2pt,qblue]([yshift=0.395cm]a57.west)--([yshift=-0.395cm]a57.west);
\draw[->,line width=2pt,qorange]([xshift=-0.395cm]a57.south)--([xshift=0.395cm]a57.south);
\draw[->,line width=2pt,qblue]([yshift=0.395cm]a67.east)--([yshift=-0.395cm]a67.east);
\draw[->,line width=2pt,qblue]([yshift=0.395cm]a77.east)--([yshift=-0.395cm]a77.east);

\node[textonly]at([yshift=0.2cm]a11.north){$s_1$};
\node[textonly]at([yshift=0.2cm]a12.north){$s_2$};
\node[textonly]at([yshift=0.2cm]a13.north){$s_3$};
\node[textonly]at([yshift=0.2cm]a14.north){...};

\node[textonly]at([xshift=-0.2cm]a11.west){$y_1$};
\node[textonly]at([xshift=-0.2cm]a21.west){$y_2$};
\node[textonly]at([xshift=-0.2cm]a31.west){$y_3$};
\node[textonly]at([xshift=-0.2cm]a41.west){...};

\node[textonly](sourcetext)at([yshift=1cm]a14) {Source side};

\node[textonly,rotate=270](targettext)at([xshift=-1cm]a41) {Target side};

\draw[-,thick,qorange]([xshift=-0.395cm]a71.south)--([xshift=-0.395cm,yshift=-0.45cm]a71.south);
\draw[-,thick,qorange]([xshift=-0.395cm]a73.south)--([xshift=-0.395cm,yshift=-0.45cm]a73.south);

\draw[<->,line width=2pt,qorange]([xshift=-0.395cm,yshift=-0.25cm]a71.south)--([xshift=-0.395cm,yshift=-0.25cm]a73.south);
\node[textonly]at([xshift=0.35cm,yshift=-0.4cm]a71.south){\textit{k}=2};

\draw[->,line width=2pt,qblue]([xshift=0.35cm]a17.east)--([xshift=0.35cm,yshift=0.1cm]a27.east);
\node[textonly]at([xshift=0.8cm,yshift=0.2cm]a27.south) {Write};
\draw[->,line width=2pt,qorange]([xshift=0.5cm,yshift=0.4cm]a37.south)--([xshift=1.1cm,yshift=0.4cm]a37.south);
\node[textonly]at([xshift=0.4cm,yshift=0.4cm]a47.east) {Read};

\end{tikzpicture}
\caption{Wait-\textit{k} policy. The model first waits for \textit{k} units (here \textit{k}=2) and then emits target word $y_t$ given source units $s_1 ... s_{t+k-1}$.}
\label{waitk}
\end{figure}
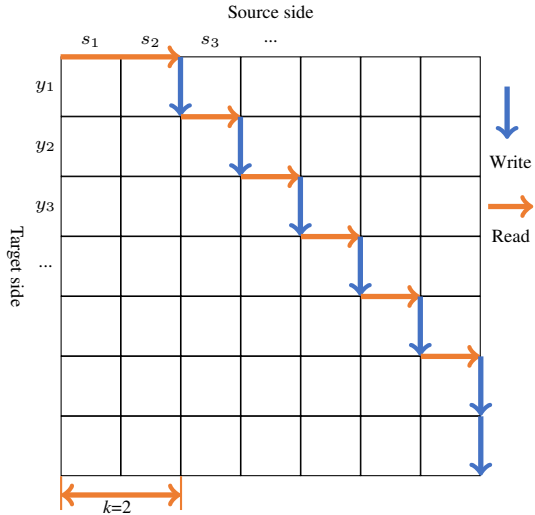

\section{Simultaneous Read-Write Policies}
\label{sec:3}
Simul R-W policies aim to identify suitable moments for generating partial sentence translations based on streaming speech units. In section \ref{sec:3.1}, we first introduce the fixed R-W policy wait-\textit{k} and its variants. Subsequently, in Section \ref{flexible}, we delve into flexible R-W policies. This section is further divided into attention-based encoder-decoder models (Section 3.2.1) and transducer-based models (Section 3.2.2) based on the model architectures. With offline ST already making significant strides, an increasing number of studies are exploring methods to render offline models in real-time, a topic we discuss in Section 3.2.3.

\subsection{The Wait-\textit{k} Method and its Variants}
\label{sec:3.1}
In the process of SimulST, simultaneous decoding policies play a pivotal role, with wait-\textit{k} \cite{waitk} emerging as a fundamental and wide approach in the early stages of simultaneous MT tasks. The underlying concept of wait-\textit{k} is to momentarily defer the output of translation until the model has received the \textit{k} units of the source input, to accumulate more context information and enhance translation accuracy.

As shown in Figure \ref{waitk}, when streaming speech input begins, the model waits for \textit{k} source units and alternates back and forth between Wait($\cdot$) and Read($\cdot$). This policy can reduce translation errors to a certain extent, because the model waits for enough contextual information before translation, to have a better understanding \cite{metalearn}. On this basis, according to the number of read and write units per time, two variants are proposed. \citet{realtrans} propose wait-\textit{k} stride-\textit{n}, in which the model alternates between writing and reading every \textit{n} units instead of \textit{one}. Besides, \citet{waitkstriden} propose wait-(\textit{k, s, n}), that is generating \textit{n} units after reading  \textit{s} additional units. They are both wait-\textit{k} policies with inconsistent step sizes.

As a straightforward approach, this simple way of simul R-W policy is easy to implement with acceptable quality and has been adopted by numerous studies \cite{SimulSpeech,augmemory,ImplicitMemory,Shiftable}. However, as it fixed alternates in reading and writing without analyzing the currently received input units, it cannot clarify whether the present moment is suitable for output. Therefore, it is often employed in conjunction with segmentation strategies in Section \ref{seg}.

\input{fig/fig4}

\subsection{Flexible Policies}
\label{flexible}

Intuitively, fixed policies like wait-\textit{k} may not have sufficient information to generate tokens based on partial inputs. In principle, model-based policies should be capable of adaptation by considering potential alignments between input segments and output tokens during training \cite{SimulMTtoSimulST}. Two types of fundamental frameworks are depicted in Figure \ref{models}: (a) indicates the Attention-based Encoder-Decoder (AED) structure which allows the decoder to attend to a portion of the source sequence without being constrained by specific modes or sequences, while Transducer in (b) is known for its advantages of monotonic alignment capability.

\subsubsection{Attention-based Encoder-Decoder Models}
\label{sec:3.2.1}

Building upon the advantage of being able to flexibly attend to relevant parts of the input, some studies tend to incorporate monotonic capability into AED models. \citet{SimulMTtoSimulST} extend Monotonic Multi-head Attention (MMA) to SimulST for achieving flexible decision-making. MMA achieves flexible decision-making by assigning each head within a layer an independent step probability, which determines when to read or write during the translation process. It is more robust to the granularity of the input, and a pre-decision module is introduced to handle fine-grained input. Functionally, the pre-decision is consistent with the segmentation strategies in Section \ref{seg}, aiming to achieve suitable segmentation units. Furthermore, \citet{attentive} propose Decision Attentive Regularization (DAR) to improve SimulST by implicitly utilizing the monotonic attention energies seen in SimulMT. 


Like CTC or MMA, Continuous Integrate-and-Fire (CIF) is another monotonic alignment method proposed to learn the precise acoustic boundaries \cite{whentotrans}. Thus, \citet{cif} leverage CIF to develop a flexible policy. Specifically, they utilize a weight prediction network and establish a threshold. If the accumulated weights fall below the threshold, CIF proceeds to the next encoder step. Otherwise, it triggers the integrate and fire operation, which retains the remaining weight for the next integration and produces an integrated embedding sent to the decoder, a process referred to as firing.

\definecolor{qblue}{RGB}{68,114,196}
\definecolor{qpurple}{RGB}{187,161,203}
\definecolor{qgreen}{RGB}{112,173,71}
\definecolor{qorange}{RGB}{237,125,49}
\definecolor{qpink}{RGB}{249,199,207}
\definecolor{qred}{RGB}{255,99,71}
\definecolor{qyellow}{RGB}{255,192,0}
\definecolor{sgreen}{RGB}{197,224,180}
\definecolor{syellow}{RGB}{255,230,153}
\definecolor{sblack}{RGB}{242,242,242}
\definecolor{backyellow}{RGB}{251,232,218}
\definecolor{backgreen}{RGB}{229,241,221}
\definecolor{backblue}{RGB}{189,215,238}
\begin{figure}
\centering
\begin{tikzpicture}
    \tikzstyle{textonly} = [font=\scriptsize,align=left]
    \tikzstyle{sublayer} = [rectangle,draw,minimum width=1.7cm,rounded corners=3pt,align=center,inner sep=3pt,minimum height=0.9cm,font=\scriptsize];
    \tikzstyle{squrd} = [rectangle,draw,minimum width=0.4cm,rounded corners=3pt,align=center,inner sep=3pt,minimum height=0.4cm,font=\scriptsize];
    \tikzstyle{middlesqurd} = [rectangle,draw,minimum width=0.3cm,rounded corners=2pt,align=center,inner sep=3pt,minimum height=0.3cm,font=\scriptsize];
    \tikzstyle{smallsqurd} = [rectangle,draw,minimum width=0.2cm,rounded corners=1pt,align=center,inner sep=3pt,minimum height=0.2cm,font=\scriptsize];
    \tikzstyle{mycircle} = [circle,draw,minimum width=0.3cm,rounded corners=3pt,align=center,inner sep=3pt,minimum height=0.3cm,font=\scriptsize];
    \tikzstyle{mincircle} = [circle,draw,minimum width=0.25cm,rounded corners=3pt,align=center,inner sep=0pt,minimum height=0.25cm,font=\scriptsize];
    \tikzstyle{background} = [rectangle,draw,rounded corners=3pt,minimum width=2.2cm,minimum height=1.5cm,font=\footnotesize,align=center];
    \tikzstyle{back} = [rectangle,rounded corners=3pt,minimum width=5cm,minimum height=1.8cm,font=\footnotesize,align=center];

    \node[textonly](a) at(0,0) {};

    \node[sublayer,draw=qorange,fill=white,fill opacity=1](aed) at ([xshift=1.1cm,yshift=1.5cm]smma.north) {AED Encoder};
    \node[sublayer,draw=qblue,fill=white,fill opacity=1](joiner) at ([yshift=1.1cm,xshift=-1cm]aed.north) {Joiner \\  {w/o self-attn}};
    \node[sublayer,draw=qgreen,fill=white](predictor) at ([yshift=1.1cm,xshift=1cm]aed.north) {Predictor \\  {w/o cross-attn}};
    \node[textonly](s1)at([yshift=-0.6cm]aed.south){Speech};
    \node[textonly](tra)at([yshift=0.5cm]joiner.north){Translation};
    \node[textonly]at([yshift=-1.2cm]aed.south){(a) CAAT};
    \draw[->,thick](s1.north)--(aed.south);
    \draw[->,thick](aed.north)--([yshift=0.3cm]aed.north)--([yshift=0.3cm,xshift=-1cm]aed.north)--(joiner.south);
    \draw[->,thick](predictor.west)--(joiner.east);
    \draw[->,thick](joiner.north)--(tra.south);

    \node[textonly](strans) at (2,0) {};
    \node[sublayer,draw=qorange,fill=white,fill opacity=1](shared) at ([xshift=3.1cm]aed.east) {Shared \\ AED Encoder};
    \node[sublayer,draw=qblue,fill=white,fill opacity=1](joiner2) at ([yshift=1.1cm,xshift=-1cm]shared.north) {Joiner};
    \node[sublayer,draw=qgreen,fill=white](predictor2) at ([yshift=1.1cm,xshift=1cm]shared.north) {AED Decoder/\\ Predictor};
    \node[textonly](s2)at([yshift=-0.6cm]shared.south){Speech};
    \node[textonly](rloss)at([yshift=0.5cm]joiner2.north){Translation};
    \node[textonly]at([yshift=-1.2cm]shared.south){(b) TAED};
    \draw[->,thick](s2.north)--(shared.south);
    \draw[->,thick](shared.north)--([yshift=0.3cm]shared.north)--([yshift=0.3cm,xshift=-1cm]shared.north)--(joiner2.south);
    \draw[->,thick,red](shared.north)--([yshift=0.3cm]shared.north)--([yshift=0.3cm,xshift=1cm]shared.north)--(predictor2.south);
    \draw[->,thick](predictor2.west)--(joiner2.east);
    \draw[->,thick](joiner2.north)--(rloss.south);

\end{tikzpicture}
\caption{Structures combining AED and Transducer in 3.2. (a) is CAAT, which divides the Transformer decoder like the Transducer. (b) is TAED, providing a hybrid of the two models.}
\label{caatandtaed}
\end{figure}
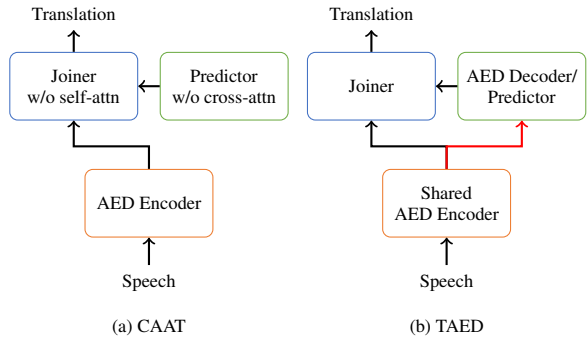

\subsubsection{Transducer Models}
\label{sec:transducer}
Another widely-used framework is RNN Transducer (RNN-T) \cite{rnn-t}. As a variant of CTC, RNN-T divides the decoder into a \textit{predictor} and a \textit{joiner}, where the predictor generates historical representations, and the joiner generates output probabilities by combining the representations of the predictor and the encoder. RNN-T naturally handles the monotonic alignment between input and output sequences during streaming decoding. Following RNN-T architecture, \citet{improstab} propose a revision-controllable method to improve the decoding stability for those utilizing beam search. 

Building upon RNN-T's foundational concepts, the Cross Attention Augmented Transducer (CAAT) \cite{caat} reconfigures the Transformer decoder into two distinct components: a predictor and a joiner. Both modules retain the original count of Transformer blocks; however, the predictor omits the cross-attention mechanism, and conversely, the joiner lacks a self-attention component as shown in Figure \ref{caatandtaed} (a).
In this way, CAAT realizes the separation of goal and historical representation in different attention mechanisms, realizing simultaneous generation. 
Further evolving this landscape, \citet{taed} introduce the TAED model, a novel hybrid of the Transducer and Attention Encoder-Decoder (AED) frameworks in Figure \ref{caatandtaed} (b). TAED utilizes a shared encoder across both paradigms while substituting the traditional Transducer predictor with an AED decoder. This strategic integration harnesses the respective advantages of AED and Transducer models, offering a unified and potent approach.

\subsubsection{Offline-to-Simul}
\label{sec:3.2.3}
Some studies focus on leveraging well-trained offline models in SimulST tasks. Regardless of R-W policies, SimulST models are typically trained only when partial input is available. However, offline models have access to entire speech during training, leading to a discrepancy when employing offline-trained models for simultaneous inference. 

To incorporate simultaneous settings to achieve incremental inference with offline models, \citet{lowlatency} aim to trade some latency for better output quality. They propose three techniques of partial hypothesis selection, which observes the acoustic information and selectively outputs stable segment-level hypotheses instead of all predictions. A further study \cite{increblockwise} propose an incremental blockwise beam-search (IBWBS) algorithm. IBWBS halts only the problematic beam upon detecting an unreliable hypothesis, allowing other beams to continue. This enables incremental SimulST and facilitates latency control without retraining the model.

Compared with adding simultaneous modules, \citet{doessimul} question the need for simultaneous training. They adopt the wait-\textit{k} policy to an offline ST model only at inference time without any additional training or adaptation. It reduces the computational costs for training a SimulST from scratch without performance degradation. 

Since \textit{k} policy is simple, they then continue to make unremitting efforts in the process of real-time adaptation of offline models. Different from the flexible policies in Section \ref{flexible} which need dedicated simulating training, they propose Encoder-Decoder Attention (EDATT) \cite{attnasaguide}. It guides offline ST models during simultaneous translation by leveraging the encoder-decoder attention matrix. If the attention is not focused on the most recent frames, the model determines to emit a partial translation since the received information is sufficient. Furthermore,  they assume that if a candidate token is aligned with the last input frame, the information might be insufficient for emission. So they present ALIGNATT \cite{AlignAtt} to guide an offline ST model during simultaneous inference by leveraging speech-to-translation alignments computed from the attention weights.

That is to say, simul R-W policies endeavor to ascertain the timing of token generation during inference, as shown in Figure \ref{overview} \textit{(b)}.
Consequently, a new trend is to leverage offline ST models during simultaneous inference, which can not only harness superior performance but also mitigate computational resource consumption.

\begin{table*}[t!]
    \centering
    \begin{threeparttable}[b]
    \footnotesize
    \begin{tabular}{lcccc}
    \toprule
    Domain &  Language & Avg. duration(h) & Avg.SacreBLEU & Common Datasets\\
    \midrule
    \multirow{3}*{ Audiobook }  & en $\rightarrow$ fr & 236 & \multirow{3}*{19.4}  & \multirow{3}*{ LibriTrans} \\
     & de $\rightarrow$ en & 100 &  \\
     & en $\rightarrow$ de & 53  &  \\
    \specialrule{0em}{1pt}{1pt}
    \cdashline{1-5}
    \specialrule{0em}{1pt}{1pt}     
    \multirow{2}*{ Lecture } & de $\rightarrow$ en & 37 & \multirow{2}*{24.4} & \multirow{2}*{ BSTC }\\
     & zh $\rightarrow$ en & 51 &  &   \\
    \specialrule{0em}{1pt}{1pt}
    \cdashline{1-5}
    \specialrule{0em}{1pt}{1pt}   
    \multirow{1}*{Common Voice} & \makecell{en  $\rightarrow$ \{fr, de, es, ca, it, ru, zh, pt, fa, et, mn, nl,\\ tr, ar, sv, lv, sl, ta, ja, if, cy \} }& 568 & 21.88 & CoVoST2\\
    \specialrule{0em}{1pt}{1pt}
    \cdashline{1-5}
    \specialrule{0em}{1pt}{1pt}  
\multirow{3}*{ TED } &  en $\rightarrow$ de & 272 & \multirow{3}*{30.4}  & \multirow{3}*{ MuST-C }\\
    & en $\rightarrow$ zh & 542 &  \\
    & en $\rightarrow$ \{ar, cs, de, es, fa, fr, it, nl, pt, ro, ru, tr, vi, zh\} & 430 &  &  \\
    \bottomrule
    \end{tabular}
    \caption{ST datasets across various domains.}
    \label{stdata}

  \end{threeparttable}
\end{table*}

\section{Evaluation Metrics}
\label{sec:4}

Achieving a balance between translation quality and latency is crucial in the SimulST task, ensuring a satisfactory user experience by providing timely translations without compromising accuracy. Introducing multiple evaluation metrics is important for assessing the balance between quality and latency comprehensively. Different metrics offer diverse perspectives on translation performance, enabling a more nuanced understanding of system behavior. By considering both quality-related and latency-related metrics, researchers can make informed decisions about system optimizations.

\subsection{Quality-based Metrics} 

The quality evaluation metrics utilized in SimulST are fundamentally aligned with those employed in MT, as both aim to assess the quality of the translated output. BLEU \cite{bleu} is one of the most commonly used metrics for evaluating the quality of MT based on measuring the closeness of a machine translation to human reference. The score ranges from 0 to 1, with higher scores indicating better translation quality. 
SacreBLEU \cite{sacrebleu} is an improved version of BLEU that offers additional features like handling tokenization issues, supporting multiple languages, and providing more robust evaluation across different datasets. 
Another recent metric is COMET \cite{comet}, aiming to capture not only surface-level similarities but also deeper semantic and contextual aspects of translations. It incorporates multiple sub-metrics, including fluency, adequacy, and fidelity, to offer a holistic assessment of translation quality. 

\subsection{Latency-based Metrics} 

The evaluation of latency metrics in SimulST serves to gauge real-time system performance. Average Lagging (AL) \cite{waitk} refers to the measure of average delay. 
It is typically calculated as the average time taken from the arrival of a speech segment to the completion of translation for that segment. 
In SimulST, the input features come from the speech $S = \{s_1, ..., s_{|S|}\}$, in which $s_i$ is a raw audio segment of duration $T_i$. Assuming ${\hat{s}}$ has been read to  generate $y_t$, the delay of $y_t$ can be defined as $d_t = \sum_{j=1}^{g(t)}T_j$. 
For a Simul R-W policy $g$, it calculates the average delay from the generation of the first target token to the $\tau(|s|)$-th, which can be defined as:
\begin{eqnarray}
AL = \frac{1}{\tau(|S|)} \sum_{t=1}^{\tau(|S|)}d_t - \frac{(t - 1)}{r}
\end{eqnarray}
where $\tau(|S|) = min\{t|d_t = |S|\}$ denotes the truncation step of the policy and $r = |Y|/|S|$ denotes the ratio between the target and source length. It can be inferred that $(t-1)/r$ term is the ideal policy to compare.

 Considering the speech duration $T$, \citet{simuleval} adapt AL as follows:
\begin{eqnarray}
AL = \frac{1}{\tau'(|S|)} \sum_{t=1}^{\tau'(|S|)}d_t - d^*_t
\end{eqnarray}
where $\tau'(|S|) = min\{t|d_t = \sum_{j=1}^{|S|}T_j\}$ and the $d^*_t = (t-1)\sum_{j=1}^{|S|}T_j/|Y_{ref}|$ are the delays of an ideal policy. It assumed that the ideal policy generates the reference $Y_{ref}$ rather than the system hypothesis and is a wait-0 policy.

Differentiable Average Lagging (DAL) \cite{dal} is computed similarly to AL but incorporates differentiable operations, allowing the latency metric to be optimized alongside the training process of the model. 
\begin{eqnarray}
\label{dal}
DAL = \frac{1}{|Y|} \sum_{t=1}^{|Y|}d'_t - \frac{(t - 1) }{r} 
\end{eqnarray}
where 
\begin{eqnarray}
d'_t = \left.
\begin{cases} 
d_t & {t = 1}\\
max[d_t,d'_{t-1} + r] & {t > 1}
\end{cases} 
\right.
\end{eqnarray}
$d'_t$ tracks duration before generating $y_t$, reflecting the semantics of $d_t$. The recursion in $d'_t$ is differentiable and can be efficiently implemented in computational graph-based programming languages.

Besides, \citet{SimulMTtoSimulST} introduce a computation-aware (CA) and a non-computation-aware (NCA) delay adapted from AL and \citet{atd} propose Average Token Delay (ATD) that focuses on the end timings of partial translations in SimulST. These various metrics are all used to evaluate latency, which needs a toolkit to apply. SimulEval \cite{simuleval} uses a server-client scheme to imitate simultaneous scenarios, it automatically performs streaming decoding and collectively reports several popular latency metrics.

Under reasonable experimental settings, subjective evaluation often has more accurate performance. Unlike the evaluation indicators mentioned above, Continuous Rating (CR) \cite{cr} is a method for human assessment of SimulST quality. It aims to provide a comprehensive assessment of the system's real-time responsiveness and translation quality. CR measures user satisfaction with the system's outputs through continuous ratings, reflecting users' real-time experiences and satisfaction levels with each translation result. This evaluation method helps assess the performance in dynamic and real-time interactions and provides intuitive feedback to guide system improvements and optimizations.

As shown in Table \ref{stdata}, we present several commonly used speech datasets (like LibriTrans\footnote{\href{https://www.openslr.org/12/}{https://www.openslr.org/12/}}, BSTC\footnote{\href{https://aistudio.baidu.com/competition/detail/44}{https://aistudio.baidu.com/competition/detail/44}}, CoVoST2\footnote{\href{https://huggingface.co/datasets/covost2}{https://huggingface.co/datasets/covost2}}, MuST-C\footnote{\href{https://mt.fbk.eu/must-c/}{https://mt.fbk.eu/must-c/}}) across different domains, along with their average SacreBLEU scores. Specifically, while models are often trained on multiple datasets, evaluations using the AL metric are typically conducted on the MuST-C corpus, which is the most commonly used in ST, ensuring comparability across different studies.
\section{Augmented Training Methods} 
\label{sec:5}

The data scale varies significantly across ASR, MT, and ST tasks, with ST datasets notably smaller due to high annotation costs. As shown in Table \ref{datascale}, for a specific language pair, the training data for speech translation tasks typically consists of only a few hundred hours, while the training data for related ASR and MT tasks exceeds it by nearly a hundredfold. Section 5.1 discusses data augmentation methods, while Section 5.2 explores studies based on the multi-task learning framework.

\begin{table}[t!]
    \centering
    \footnotesize
    \begin{tabular}{ccc}
    \toprule
    Task &  Modeling & Data Scale \\
    \midrule
    ASR  & Cross-modal & 100K hours \\
    MT & Cross-lingual & 1B sentences \\
    SimuST & Cross-modal and cross-lingual & Ks hours \\
    \bottomrule
    \end{tabular}
    \caption{The scale of annotated data for ASR, MT, and ST tasks for a specific language pair (like en $\rightarrow$ de).}
    \label{datascale}
\end{table}

\subsection{Data Augmentation}
Training SimulST models, which rely on ST data, presents a significant challenge due to data scarcity. Despite ST data is not specifically tailored for streaming tasks like SimulST, segmentation tools can be employed to adapt the data, they also pose challenges such as model convergence difficulties and lack of robustness. Data augmentation is an effective means to expand training data. For effective training, \citet{xiaomi} utilize a well-trained MT model to translate the transcriptions from ASR data and synthesize a large amount of pseudo-data. While expanding data, they also used pre-training model weights to initialize SimulST, to make use of the pre-trained models. Rather than a simple data mixture, \citet{Tagged} propose a method to address the scarcity of simultaneous interpretation (SI) data by using a larger-scale offline translation corpus for training. They also introduce a tag-based approach to control the style of translations and handle the differences.

\subsection{Multi-task Learning}

Multi-task learning is an enhancing approach to model training because it allows for learning multiple related tasks simultaneously, providing additional information and constraints to improve model performance. \citet{lowlatencymtl} investigate two methods to select reasonable sub-strings from the reference to build partial parallel corpora for model-training and they opt to use multi-task learning to take advantage of a pre-trained MT model. A similar idea brings Modality Agnostic Meta-Learning (MAML) in SimulST involving meta-learning and fine-tuning steps \cite{metalearn}. In the former step, a set of high-resource tasks are trained as source tasks to capture general aspects; then they fine-tune the model further to learn the specific SimulST task in the latter step. Except for DAR, \citet{attentive} employ multi-task learning by training the SimulMT model along with SimulST and \citet{SH} use multi-task learning to jointly learn tasks with a shared encoder, leveraging streaming ASR to guide SimulST decoding via beam search. 
Based on AED, \citet{blockwisetrans} achieve a joint CTC/attention by injecting a CTC objective between encoder outputs and target translations. They additionally calculate a CTC loss from the CTC branch and apply an ASR-based intermediate CTC loss for multi-task learning.

By jointly training multiple tasks within the same model, it enables the sharing of underlying representations, thereby improving data efficiency. As augmented training methods, when combining data enhancement and multi-task learning methods with the original, the model can benefit more as well as be fully trained.

\section{Future Work}
\label{sec:6}

Based on the recent advancements in SimulST tasks and the demands of real-time scenarios, we believe that two promising directions are multilingual SimulST and integration with Large Language Models.

\subsection{Multilingual SimulST}

With the advancement of globalization and increased cross-cultural communication, multilingual SimulST holds significant potential. This approach enables real-time translation of speech inputs into multiple languages, facilitating communication and collaboration in multilingual environments. 

Following the simultaneous adaptation procedure in the previous work \cite{lowlatency}, \citet{Multilingual} explore whether it can be utilized to build multilingual SimulST and they conduct experiments on both cascaded and end-to-end offline models. With the focus on multilingual SimulST, \citet{jointtraining} propose a separate decoder model and a unified encoder-decoder model for joint training and decoding. They also introduce an asynchronous training strategy to enhance knowledge transfer among different languages. Based on the neural transducer, \citet{wang2022lamassu} introduce LAMASSU for language-agnostic multilingual speech recognition and translation in a streaming fashion. By incorporating multilingual capabilities, the system becomes more versatile, accommodating diverse language settings and catering to the needs of various demographics.

\subsection{Integration with LLMs}
In recent years, the development of large-scale language models has made significant strides \cite{zhao2023survey}.
Large language models (LLMs) leverage extensive pre-existing linguistic knowledge \cite{whisper}, thereby improving translation quality and accuracy \cite{ComSL}. Among this, AudioPaLM combines text-based and speech-based language models into a multi-modal generative model, covering most of the offline text and speech processing and generation tasks \cite{AudioPaLM}. A brand-new work Seamless releases a set of full-process large-scale speech translation systems, introducing a family of Seamless models that enable multilingual and expressive SimulST \cite{seamless}. Integrating LLMs into SimulST systems enhances their ability to accurately understand speech inputs, handle contextual dependencies, and generate fluent translations. In a word, we anticipate that we can further enhance the performance and applicability of streaming speech translation systems, meeting the diverse needs of users in real-time scenarios by combining these two directions.

\section*{Acknowledgements}

This work was supported in part by the National Science Foundation of China (No.62276056), the Natural Science Foundation of Liaoning Province of China (2022-KF-16-01), the Fundamental Research Funds for the Central Universities (Nos. N2216016 and N2316002), the Yunnan Fundamental Research Projects (No. 202401BC070021), and the Program of Introducing Talents of Discipline to Universities, Plan 111 (No.B16009).

\section*{Contribution Statement}

\noindent \textbf{Xiaoqian Liu} and \textbf{Guoqiang Hu} conducted the literature search. They collected and categorized the relevant papers, and drafted the main sections of the manuscript. They also integrated the feedback from other contributors to refine and enhance the overall quality of the manuscript.

\textbf{Yangfan Du}, \textbf{Erfeng He}, and \textbf{Yingfeng Luo} were responsible for the creation of formulas, tables, and figures, and they also provided suggestions to ensure the accuracy and coherence of the paper. 


\textbf{Chen Xu}, \textbf{Tong Xiao}, and \textbf{Jingbo Zhu} planned and guided the overall direction of the survey. They provided substantial feedback at various stages of the manuscript and offered insights on future trends in SimulST.


\bibliographystyle{named}
\bibliography{ijcai24}

\begin{thebibliography}{}

\bibitem[\protect\citeauthoryear{Chang and Lee}{2022}]{cif}
Chih{-}Chiang Chang and Hung{-}yi Lee.
\newblock Exploring continuous integrate-and-fire for adaptive simultaneous speech translation.
\newblock In {\em Interspeech}, 2022.

\bibitem[\protect\citeauthoryear{Chen \bgroup \em et al.\egroup }{2021}]{SH}
Junkun Chen, Mingbo Ma, Renjie Zheng, and Liang Huang.
\newblock Direct simultaneous speech-to-text translation assisted by synchronized streaming {ASR}.
\newblock In {\em ACL Findings}, 2021.

\bibitem[\protect\citeauthoryear{Chen \bgroup \em et al.\egroup }{2023}]{improstab}
Junkun Chen, Jian Xue, Peidong Wang, Jing Pan, and Jinyu Li.
\newblock Improving stability in simultaneous speech translation: {A} revision-controllable decoding approach.
\newblock {\em CoRR}, 2023.

\bibitem[\protect\citeauthoryear{Cherry and Foster}{2019}]{dal}
Colin Cherry and George~F. Foster.
\newblock Thinking slow about latency evaluation for simultaneous machine translation.
\newblock {\em CoRR}, 2019.

\bibitem[\protect\citeauthoryear{Communication \bgroup \em et al.\egroup }{2023}]{seamless}
Seamless Communication, Loïc Barrault, Yu-An Chung, Mariano~Coria Meglioli, David Dale, Ning Dong, Mark Duppenthaler, Paul-Ambroise Duquenne, Brian Ellis, Hady Elsahar, et~al.
\newblock Seamless: Multilingual expressive and streaming speech translation.
\newblock {\em CoRR}, 2023.

\bibitem[\protect\citeauthoryear{Deng \bgroup \em et al.\egroup }{2022}]{blockwisetrans}
Keqi Deng, Shinji Watanabe, Jiatong Shi, and Siddhant Arora.
\newblock Blockwise streaming transformer for spoken language understanding and simultaneous speech translation.
\newblock In {\em Interspeech}, 2022.

\bibitem[\protect\citeauthoryear{Dong \bgroup \em et al.\egroup }{2022}]{whentotrans}
Qian Dong, Yaoming Zhu, Mingxuan Wang, and Lei Li.
\newblock Learning when to translate for streaming speech.
\newblock In {\em ACL}, 2022.

\bibitem[\protect\citeauthoryear{Graves \bgroup \em et al.\egroup }{2006}]{ctc}
Alex Graves, Santiago Fern{\'{a}}ndez, Faustino~J. Gomez, and J{\"{u}}rgen Schmidhuber.
\newblock Connectionist temporal classification: labelling unsegmented sequence data with recurrent neural networks.
\newblock In {\em ICML}, 2006.

\bibitem[\protect\citeauthoryear{Graves}{2012}]{rnn-t}
Alex Graves.
\newblock Sequence transduction with recurrent neural networks.
\newblock {\em CoRR}, 2012.

\bibitem[\protect\citeauthoryear{Han \bgroup \em et al.\egroup }{2020}]{metalearn}
Hou~Jeung Han, Mohd~Abbas Zaidi, Sathish~Reddy Indurthi, Nikhil~Kumar Lakumarapu, Beomseok Lee, and Sangha Kim.
\newblock End-to-end simultaneous translation system for {IWSLT}2020 using modality agnostic meta-learning.
\newblock In {\em IWSLT}, 2020.

\bibitem[\protect\citeauthoryear{Huang \bgroup \em et al.\egroup }{2023a}]{jointtraining}
Wuwei Huang, Renren Jin, Wen Zhang, Jian Luan, Bin Wang, and Deyi Xiong.
\newblock Joint training and decoding for multilingual end-to-end simultaneous speech translation.
\newblock In {\em ICASSP}, 2023.

\bibitem[\protect\citeauthoryear{Huang \bgroup \em et al.\egroup }{2023b}]{xiaomi}
Wuwei Huang, Mengge Liu, Xiang Li, Yanzhi Tian, Fengyu Yang, Wen Zhang, Jian Luan, Bin Wang, Yuhang Guo, and Jinsong Su.
\newblock The xiaomi {AI} lab's speech translation systems for {IWSLT} 2023 offline task, simultaneous task and speech-to-speech task.
\newblock In {\em IWSLT}, 2023.

\bibitem[\protect\citeauthoryear{Kano \bgroup \em et al.\egroup }{2022}]{atd}
Yasumasa Kano, Katsuhito Sudoh, and Satoshi Nakamura.
\newblock Average token delay: {A} latency metric for simultaneous translation.
\newblock {\em CoRR}, 2022.

\bibitem[\protect\citeauthoryear{Ko \bgroup \em et al.\egroup }{2023}]{Tagged}
Yuka Ko, Ryo Fukuda, Yuta Nishikawa, Yasumasa Kano, Katsuhito Sudoh, and Satoshi Nakamura.
\newblock Tagged end-to-end simultaneous speech translation training using simultaneous interpretation data.
\newblock In {\em IWSLT}, 2023.

\bibitem[\protect\citeauthoryear{Le \bgroup \em et al.\egroup }{2023}]{ComSL}
Chenyang Le, Yao Qian, Long Zhou, Shujie Liu, Michael Zeng, and Xuedong Huang.
\newblock Comsl: {A} composite speech-language model for end-to-end speech-to-text translation.
\newblock In {\em NeurIPS}, 2023.

\bibitem[\protect\citeauthoryear{Liu \bgroup \em et al.\egroup }{2020}]{lowlatency}
Danni Liu, Gerasimos Spanakis, and Jan Niehues.
\newblock Low-latency sequence-to-sequence speech recognition and translation by partial hypothesis selection.
\newblock In {\em Interspeech}, 2020.

\bibitem[\protect\citeauthoryear{Liu \bgroup \em et al.\egroup }{2021}]{caat}
Dan Liu, Mengge Du, Xiaoxi Li, Ya~Li, and Enhong Chen.
\newblock Cross attention augmented transducer networks for simultaneous translation.
\newblock In {\em EMNLP}, 2021.

\bibitem[\protect\citeauthoryear{Ma \bgroup \em et al.\egroup }{2019}]{waitk}
Mingbo Ma, Liang Huang, Hao Xiong, Renjie Zheng, Kaibo Liu, Baigong Zheng, Chuanqiang Zhang, Zhongjun He, Hairong Liu, et~al.
\newblock {STACL}: Simultaneous translation with implicit anticipation and controllable latency using prefix-to-prefix framework.
\newblock In {\em ACL}, 2019.

\bibitem[\protect\citeauthoryear{Ma \bgroup \em et al.\egroup }{2020a}]{simuleval}
Xutai Ma, Mohammad~Javad Dousti, Changhan Wang, Jiatao Gu, and Juan Pino.
\newblock {SIMULEVAL}: An evaluation toolkit for simultaneous translation.
\newblock In {\em EMNLP}, 2020.

\bibitem[\protect\citeauthoryear{Ma \bgroup \em et al.\egroup }{2020b}]{SimulMTtoSimulST}
Xutai Ma, Juan Pino, and Philipp Koehn.
\newblock {S}imul{MT} to {S}imul{ST}: Adapting simultaneous text translation to end-to-end simultaneous speech translation.
\newblock In {\em AACL}, 2020.

\bibitem[\protect\citeauthoryear{Ma \bgroup \em et al.\egroup }{2021}]{augmemory}
Xutai Ma, Yongqiang Wang, Mohammad~Javad Dousti, Philipp Koehn, and Juan~Miguel Pino.
\newblock Streaming simultaneous speech translation with augmented memory transformer.
\newblock In {\em ICASSP}, 2021.

\bibitem[\protect\citeauthoryear{Mach{\'{a}}cek \bgroup \em et al.\egroup }{2023}]{cr}
Dominik Mach{\'{a}}cek, Ondrej Bojar, and Raj Dabre.
\newblock {MT} metrics correlate with human ratings of simultaneous speech translation.
\newblock In {\em IWSLT}, 2023.

\bibitem[\protect\citeauthoryear{Nguyen \bgroup \em et al.\egroup }{2021a}]{waitkstriden}
Ha~Nguyen, Yannick Est{\`{e}}ve, and Laurent Besacier.
\newblock An empirical study of end-to-end simultaneous speech translation decoding strategies.
\newblock In {\em ICASSP}, 2021.

\bibitem[\protect\citeauthoryear{Nguyen \bgroup \em et al.\egroup }{2021b}]{impact}
Ha~Nguyen, Yannick Est{\`{e}}ve, and Laurent Besacier.
\newblock Impact of encoding and segmentation strategies on end-to-end simultaneous speech translation.
\newblock In {\em Interspeech}, 2021.

\bibitem[\protect\citeauthoryear{Niehues \bgroup \em et al.\egroup }{2018}]{lowlatencymtl}
Jan Niehues, Ngoc{-}Quan Pham, Thanh{-}Le Ha, Matthias Sperber, and Alex Waibel.
\newblock Low-latency neural speech translation.
\newblock In {\em Interspeech}, 2018.

\bibitem[\protect\citeauthoryear{Papi \bgroup \em et al.\egroup }{2022}]{doessimul}
Sara Papi, Marco Gaido, Matteo Negri, and Marco Turchi.
\newblock Does simultaneous speech translation need simultaneous models?
\newblock In {\em EMNLP Findings}, 2022.

\bibitem[\protect\citeauthoryear{Papi \bgroup \em et al.\egroup }{2023a}]{attnasaguide}
Sara Papi, Matteo Negri, and Marco Turchi.
\newblock Attention as a guide for simultaneous speech translation.
\newblock In {\em ACL}, 2023.

\bibitem[\protect\citeauthoryear{Papi \bgroup \em et al.\egroup }{2023b}]{AlignAtt}
Sara Papi, Marco Turchi, and Matteo Negri.
\newblock Alignatt: Using attention-based audio-translation alignments as a guide for simultaneous speech translation.
\newblock {\em CoRR}, 2023.

\bibitem[\protect\citeauthoryear{Papineni \bgroup \em et al.\egroup }{2002}]{bleu}
Kishore Papineni, Salim Roukos, Todd Ward, and Wei-Jing Zhu.
\newblock {B}leu: a method for automatic evaluation of machine translation.
\newblock In {\em ACL}, 2002.

\bibitem[\protect\citeauthoryear{Pol{\'{a}}k \bgroup \em et al.\egroup }{2023}]{increblockwise}
Peter Pol{\'{a}}k, Brian Yan, Shinji Watanabe, Alex Waibel, and Ondrej Bojar.
\newblock Incremental blockwise beam search for simultaneous speech translation with controllable quality-latency tradeoff.
\newblock {\em CoRR}, 2023.

\bibitem[\protect\citeauthoryear{Pol{\'{a}}k}{2023}]{longformsurvey}
Peter Pol{\'{a}}k.
\newblock Long-form simultaneous speech translation: Thesis proposal.
\newblock {\em CoRR}, 2023.

\bibitem[\protect\citeauthoryear{Post}{2018}]{sacrebleu}
Matt Post.
\newblock A call for clarity in reporting {BLEU} scores.
\newblock In {\em {WMT} 2018}, 2018.

\bibitem[\protect\citeauthoryear{Radford \bgroup \em et al.\egroup }{2023}]{whisper}
Alec Radford, Jong~Wook Kim, Tao Xu, Greg Brockman, Christine McLeavey, and Ilya Sutskever.
\newblock Robust speech recognition via large-scale weak supervision.
\newblock In {\em ICML}, 2023.

\bibitem[\protect\citeauthoryear{Raffel and Chen}{2023}]{ImplicitMemory}
Matthew Raffel and Lizhong Chen.
\newblock Implicit memory transformer for computationally efficient simultaneous speech translation.
\newblock In {\em ACL Findings}, 2023.

\bibitem[\protect\citeauthoryear{Raffel \bgroup \em et al.\egroup }{2023}]{Shiftable}
Matthew Raffel, Drew Penney, and Lizhong Chen.
\newblock Shiftable context: Addressing training-inference context mismatch in simultaneous speech translation.
\newblock In {\em ICML}, 2023.

\bibitem[\protect\citeauthoryear{Rei \bgroup \em et al.\egroup }{2020}]{comet}
Ricardo Rei, Craig Stewart, Ana~C Farinha, and Alon Lavie.
\newblock Unbabel{'}s participation in the {WMT}20 metrics shared task.
\newblock In {\em WMT}, 2020.

\bibitem[\protect\citeauthoryear{Ren \bgroup \em et al.\egroup }{2020}]{SimulSpeech}
Yi~Ren, Jinglin Liu, Xu~Tan, Chen Zhang, Tao Qin, Zhou Zhao, and Tie-Yan Liu.
\newblock {S}imul{S}peech: End-to-end simultaneous speech to text translation.
\newblock In {\em ACL}, 2020.

\bibitem[\protect\citeauthoryear{Rubenstein \bgroup \em et al.\egroup }{2023}]{AudioPaLM}
Paul~K. Rubenstein, Chulayuth Asawaroengchai, Duc~Dung Nguyen, Ankur Bapna, Zal{\'{a}}n Borsos, F{\'{e}}lix de~Chaumont~Quitry, Peter Chen, Dalia~El Badawy, Wei Han, Eugene Kharitonov, et~al.
\newblock Audiopalm: {A} large language model that can speak and listen.
\newblock {\em CoRR}, 2023.

\bibitem[\protect\citeauthoryear{Sethiya and Maurya}{2023}]{survey23}
Nivedita Sethiya and Chandresh~Kumar Maurya.
\newblock End-to-end speech-to-text translation: A survey.
\newblock {\em CoRR}, 2023.

\bibitem[\protect\citeauthoryear{Subramanya and Niehues}{2022}]{Multilingual}
Shashank Subramanya and Jan Niehues.
\newblock Multilingual simultaneous speech translation.
\newblock {\em CoRR}, 2022.

\bibitem[\protect\citeauthoryear{Tang \bgroup \em et al.\egroup }{2023}]{taed}
Yun Tang, AnnaY. Sun, Hirofumi Inaguma, Xinyue Chen, Ning Dong, Xutai Ma, PadenD. Tomasello, and Juan Pino.
\newblock Hybrid transducer and attention based encoder-decoder modeling for speech-to-text tasks.
\newblock {\em CoRR}, 2023.

\bibitem[\protect\citeauthoryear{Vaswani \bgroup \em et al.\egroup }{2017}]{transformer}
Ashish Vaswani, Noam Shazeer, Niki Parmar, Jakob Uszkoreit, Llion Jones, Aidan~N. Gomez, Lukasz Kaiser, and Illia Polosukhin.
\newblock Attention is all you need.
\newblock In {\em NeurIPS}, 2017.

\bibitem[\protect\citeauthoryear{Wang \bgroup \em et al.\egroup }{2022}]{wang2022lamassu}
Peidong Wang, Eric Sun, Jian Xue, Yu~Wu, Long Zhou, Yashesh Gaur, Shujie Liu, and Jinyu Li.
\newblock Lamassu: Streaming language-agnostic multilingual speech recognition and translation using neural transducers.
\newblock {\em CoRR}, 2022.

\bibitem[\protect\citeauthoryear{Xu \bgroup \em et al.\egroup }{2023}]{xusurvey}
Chen Xu, Rong Ye, Qianqian Dong, Chengqi Zhao, Tom Ko, Mingxuan Wang, Tong Xiao, and Jingbo Zhu.
\newblock Recent advances in direct speech-to-text translation.
\newblock In {\em IJCAI}, 2023.

\bibitem[\protect\citeauthoryear{Zaidi \bgroup \em et al.\egroup }{2021}]{attentive}
Mohd~Abbas Zaidi, Beomseok Lee, Nikhil~Kumar Lakumarapu, Sangha Kim, and Chanwoo Kim.
\newblock Decision attentive regularization to improve simultaneous speech translation systems.
\newblock {\em CoRR}, 2021.

\bibitem[\protect\citeauthoryear{Zeng \bgroup \em et al.\egroup }{2021}]{realtrans}
Xingshan Zeng, Liangyou Li, and Qun Liu.
\newblock {R}eal{T}ran{S}: End-to-end simultaneous speech translation with convolutional weighted-shrinking transformer.
\newblock In {\em ACL Findings}, 2021.

\bibitem[\protect\citeauthoryear{Zhang and Feng}{2023}]{Zhangacl}
Shaolei Zhang and Yang Feng.
\newblock End-to-end simultaneous speech translation with differentiable segmentation.
\newblock In {\em ACL Findings}, 2023.

\bibitem[\protect\citeauthoryear{Zhang \bgroup \em et al.\egroup }{2022}]{adaptiveseg}
Ruiqing Zhang, Zhongjun He, Hua Wu, and Haifeng Wang.
\newblock Learning adaptive segmentation policy for end-to-end simultaneous translation.
\newblock In {\em ACL}, 2022.

\bibitem[\protect\citeauthoryear{Zhao \bgroup \em et al.\egroup }{2023}]{zhao2023survey}
Wayne~Xin Zhao, Kun Zhou, Junyi Li, Tianyi Tang, Xiaolei Wang, Yupeng Hou, Yingqian Min, Beichen Zhang, Junjie Zhang, Zican Dong, et~al.
\newblock A survey of large language models.
\newblock {\em CoRR}, 2023.

\end{thebibliography}

\end{document}